# The computational evidence for the crucial role of the dipole cross-correlations in the polar glass-forming liquids

Kajetan Koperwas[1,][*], and Marian Paluch[1]

[1.] University of Silesia in Katowice, Institute of Physics, 75 Pułku Piechoty 1, 41-500 Chorzów, Poland

* Corresponding authors: kajetan.koperwas@us.edu.pl

**ABSTRACT**

In this letter, we analyze the dipole-dipole correlations obtained from the molecular dynamics simulations for strongly- and weakly-polar model liquids. As a result, we found that cross-correlations contribution to the systems' total dipole moment correlation function, which is directly measured in the dielectric spectroscopy experiment, is negligible for weakly-polar liquids. In contrast, the cross-correlations term dominates over the self-correlations one for examined strongly polar-liquid. Consequently, our studies strongly support the interpretation of the dielectric spectra nature of the glass forming liquids, recently proposed by Pabst *et al.*

Despite the fact that the Broadband Dielectric Spectroscopy (BDS) is an experimental technique commonly used to study molecular dynamics for more than one hundred years [1,2], the comprehensive understanding of the nature of obtained spectra is still missing. [3,4] Therefore, it is not surprising that the interpretation of BDS spectral shape describing the structural relaxation is a subject of the endless and hot debate within the physics of glass forming liquids. [5–7] In this context, the notable advantages of BDS, which is extremally broad range of measured relaxation times (18 decades) [8] and diversity of studied materials, [9] make the disclosure of the fundamental features of BDS relaxation spectra even more puzzling. Hence, supplementation of BDS spectra with results delivered by other experimental techniques probing the corresponding molecular motions is crucial. [9] One of the noteworthy technique is Depolarized Dynamic Light Scattering (DDLS), which, similarly to BDS, is sensitive to molecular reorientations. [10] However, the simple comparison of BDS and DDLS spectra reveals significant differences. The stretching parameter describing the dielectric-loss peak corresponding to $\alpha$ relaxation process usually varies from 0.5 to 0.9 for weak and strong polar liquids, respectively. [4] In contrast, DDLS spectra reveal quasi-universal shape of stretching parameter value approximately equal to 0.5. [11]

Recently, the promising solution to the problem, and hence the entirely new light on the origin of broadening of the relaxation spectra registered by BDS, has been put by Pabst *et al.* [12,13] Authors postulated that the spectrum detected by a BDS originates from two types of correlations occurring between molecules of polar liquids. The first one, which vanishes faster, describes how long a given molecule "remembers" its initial orientations. It is so-called self-correlation. The second one describes the time evolution of the other molecules in respect to the initial orientation of the chosen one and its so-called cross-correlation. In contrast to a BDS experiment, a DDLS probes mainly the self-correlations. Experimentally it has been shown that the spectra obtained by both methods correspond to each other only for the weakly polar liquids. This accordance was explained assuming a negligible role of the cross-correlations. However, when the polarity of molecules increases, the role of the cross-correlations presumably increases as well. Therefore, the different values of the stretching parameter for BDS and DDLS spectra are observed. Hence, one can expect that the narrower shape of the dielectric spectra results from the existence of noticeable cross-correlations. By extracting the DDLS spectrum from BDS one, Pabst *et al.* tried to reveal the nature of the cross-correlations spectral shape. Performing such analysis, they pointed out that the shape of cross-correlations spectrum can be described by the Debye function (for illustration, see Fig. 1)

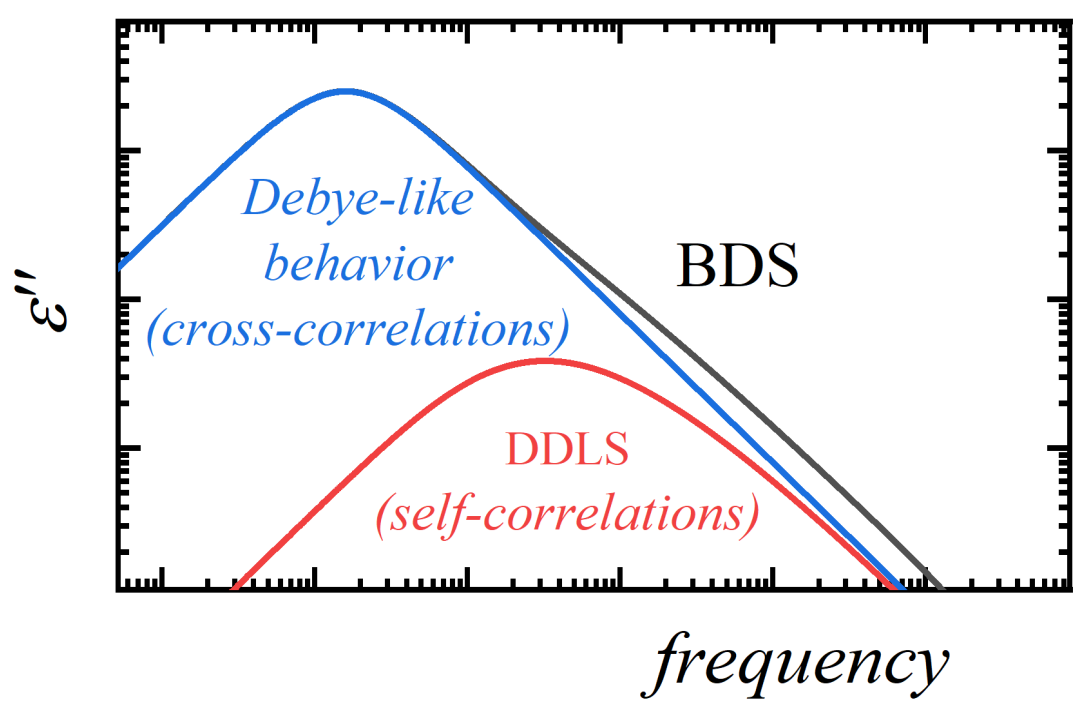

Fig.1 (color online)
The schematic representation of BDS and DDLS spectra is presented. The black line consists of two relaxation processes resulting from self- (red line) and cross- (blue line) correlations. Cross-correlations are characterized by Debye-like behavior, whereas self-correlations are described by the decay function with stretching parameter equal to 0.5.

At this point it must be noted that the common interpretation of the spectra broadening relates this phenomenon to the existence of the relaxation time distribution. [14–16] Consequently, Pabst *et al.* suggesting that the separation of time scales of the self- and cross-correlations is a main factor leading to the broader shape of the dielectric spectrum put the new light on the problem of proper interpretation of the BDS results. Moreover, their postulate is supported by the theory proposed by Déjardin *et al.* [17,18], according to which the response of a pair of dipoles consists of the two relaxation processes associated to the single and collective molecular motions. The relaxation times of those processes depend on the parameter $\lambda \sim \mu^2$ ( $\mu$ is a molecular dipole moment), increase of which separates the time scales of both processes. Hence, the experimentally observed correlation between the dielectric strength and the shape of the dielectric loss peak can be qualitatively rationalized. [13] Furthermore, the recent experimental studies devoted to the strongly polar liquid revealed that two characteristic time scales are visible during the aging process. [19] The one process mimics the generic structural relaxation toward the equilibrium, whereas the second one is ascribed to the evolution of the cross-correlations mode. Interestingly, the similar experiment performed for weakly polar liquid with negligible dipole-dipole interactions results in a recovery process exhibiting a single exponential character. However, it must be stressed that so far, there is no direct experimental evidence that the cross-correlations are marginal for weakly polar liquids. Moreover, it has not been directly shown that the cross-correlations decay slower and that their magnitude increases with the polarity of a molecule. Therefore, the proposed concept of BDS spectra nature needs to be directly verified.

In this letter, using the molecular dynamics simulations, we lay the cornerstone for any further studies based on proposed explanation of the origin of BDS spectrum shape. For the two model systems, which differ exclusively in the value of the dipole moment, we calculate the self- and cross-correlations. Subsequently, we directly analyze their shapes and estimate their relative contributions to the correlation function of the system's total dipole moment, which is probed in the BDS experiment. Our findings confirm not only that the cross-correlations dominate for polar liquids but also that they relax in the manner similar to the Debye-like behavior. Consequently, we verify the suggested interpretation of the BDS spectral shape for the first time and deliver the strong evidence for its correctness.

The dielectric experiment is designed to monitor changes in the material polarization $\boldsymbol{P}$ induced by an external electric field $E$, which changes with time. [16] However, typically it is realized in the frequency domain, which means that the periodic $E(\omega)$ is used, where $\omega$ is the angular frequency. Then the real $\epsilon'$ and the imaginary $\epsilon''$ parts of the complex dielectric permittivity, $\epsilon^*(\omega) = \epsilon'(\omega) + i\epsilon''(\omega)$, reveal an existence of the relaxation processes, i.e., around $\omega$ corresponding to the characteristic time of the given relaxation process, $\tau$, $\epsilon'(\omega)$ exhibits step-like decrease, whereas relaxation peak is detectable in $\epsilon''(\omega)$. It is due to that both parts of $\epsilon^*(\omega)$ are related to the Fourier-Laplace transformation of the decay function of polarization $C(t) = \langle \boldsymbol{P}(t)\boldsymbol{P}(0) \rangle$, $\langle \ \rangle$ denotes an ensemble average. In the simplest case of the Debye-like behavior $C(t)$ takes the exponential form, $C(t) = A exp\left(-\frac{t}{\tau}\right)$, where $A$ is a parameter. [20] However, when the spectrum exhibits broader shape, the Kohlrausch-Williams-Watts (KWW) function is commonly applied,

$$C(t) = A exp\left(-\left(\frac{t}{\tau}\right)^{\beta_{KWW}}\right), \qquad \text{Eq. (1).}$$

where $\beta_{KWW}$ is the already mentioned stretching parameter. Since $\boldsymbol{P}(t)$ is proportional the total dipole moment of the system, $\boldsymbol{M}(t) = \sum_{i=1}^{N} \boldsymbol{\mu}_i(t)$, $C(t)$ can be defined as

$$C(t) = \langle \boldsymbol{M}(0)\boldsymbol{M}(t) \rangle = \sum_{i=1}^{N}\sum_{j=1}^{N} \langle \boldsymbol{\mu}_i(0)\boldsymbol{\mu}_j(t) \rangle. \qquad \text{Eq. (2).}$$

Finally, on can rewrite $C(t)$ as a sum of the self-

$$C_s(t) = \sum_{i=1}^{N} \langle \boldsymbol{\mu}_i(0)\boldsymbol{\mu}_i(t) \rangle, \qquad \text{Eq. (3).}$$

and cross-

$$C_C(t) = \sum_{i=1}^{N}\sum_{j \neq i=1}^{N} \langle \boldsymbol{\mu}_i(0)\boldsymbol{\mu}_j(t) \rangle \qquad \text{Eq. (4).}$$

correlations terms.

According to Eq. (3) and Eq. (4), the information on the time evolution of molecules' dipole moments is needed to calculate $C_s(t)$ and $C_c(t)$. Unfortunately, neither standard experimental method measures $\boldsymbol{\mu}(t)$. However, it is easily accessible in the computer simulations of the molecular dynamics, which have another crucial advantage. Projecting model systems intended to a specific computational experiment one has entire control on the physical differences between created molecules. It is especially important because, the model systems, which are designed for the experiment targeted to the verification of the Pabst *et al.* interpretation of the BDS spectral shape, should possess identical molecular structures but differ in the value of $\mu$. Then, the model representative of weakly polar liquids would exhibit the negligible cross-correlations contribution to the total dipole correlation function, i.e., $C(t)$ should be dominated by $C_s(t)$. Contrary, for the second model system possessing substantially higher $\mu$ value, one might expect the evident role of $C_c(t)$ in $C(t)$. An attractive candidate for planned studies are the rhombus-like molecule (RM), [21–24] which consists of 4 atoms. Then the $\mu$ might be placed along one of two diagonals. It is worth mentioning that the shape of RM reflects the structural anisotropy of many Van der Walls liquids and that the RM systems are relatively easy to supercool. [23] Following our previous studies, the used atoms represent the carbon atoms, which create the benzene ring. Consistently, the bond length between RM's atoms is around $0.15nm$ (the tiny difference between real length of the bond in the benzene ring and $0.15nm$ results from the fact that one diagonal of RM is 2 times shorter than the other one). The stiffness of bonds, angles, and dihedrals, as well as the non-bonded interactions between atoms of different molecules, are defined using the parameters of OPLS all-atom force field provided for carbon atom of the benzene ring as well. [25] However, in order to create specific $\mu$ we redefine atoms' charges. In this way the studied herein RMs are characterize by $\mu$ equal to $0.386D$ and $3.86D$, which are arranged alongside the longer diagonal of the RM (this orientation impedes the crystallization of the RM systems [23]). As a starting point we equilibrate the systems containing $N = 16384$ molecules at $NVT$ conditions, $V = 1643 nm^3$ and $T = 200K$, at which $C(t)$ decays within accessible in computer simulations period of time. The Nose-Hoover thermostat [26–28], which is implemented in the GROMACS software [29–33], provided constant temperature conditions. Simulation runs were performed using velocity-Verlet integration scheme [34] with the time step equal to $0.001 ps$. The applied cutoff for intermolecular interactions was set to $4.260 nm$, which is 12 times longer than $\sigma$ parameter of intermolecular interaction potential. The long-range dipole-dipole interactions has been realized by use of the Reaction Field Method [35,36]. The dielectric constant of the reaction field, $\epsilon$, was predicted according to (the classical from the BDS point of view) Onsager theory [37], which for studied herein a non-polarizable polar molecules takes the following form $\frac{(\epsilon-1)(2\epsilon+1)}{\epsilon} = \frac{N_0 \mu^2}{\epsilon_0 k_B T}$ [38], where $N_0$ is a number of dipole within the volume unit, $\epsilon_0$ is the dielectric permittivity of vacuum, $k_B$ is a Boltzmann's constant. In Supplemental Material we show that the use of Particle Mesh Ewald summation gives an identical results [39] and that obtained results do not dependent on the system size.

As we already mentioned the BDS experiment uses the externa electric field, which is applied to the sample. Therefore, we examine RM systems not only at equilibrium conditions but also when outer disturbance is employed. Consequently, procedure of our computational experiment consists of 3 parts, which were repeated 50 times for each RM system. At the first step we polarize previously equilibrated system by an application of a constant external electric field in direction $z$, $E_z$. It has to be noted that applied $E_z$ are equal to half of $0.1 k_B T/\mu$, which ensures that our experiment is performed in a linear response regime. [16] In the next step, an applied external electric field is suddenly turn off, which enables estimation of the relaxation functions of $P$ directly probed in BDS experiment. The final step is a standard simulation without an external field for $0.2 ns$, which is carried out to calculate $C(t)$. The obtained results are presented in Fig 2.

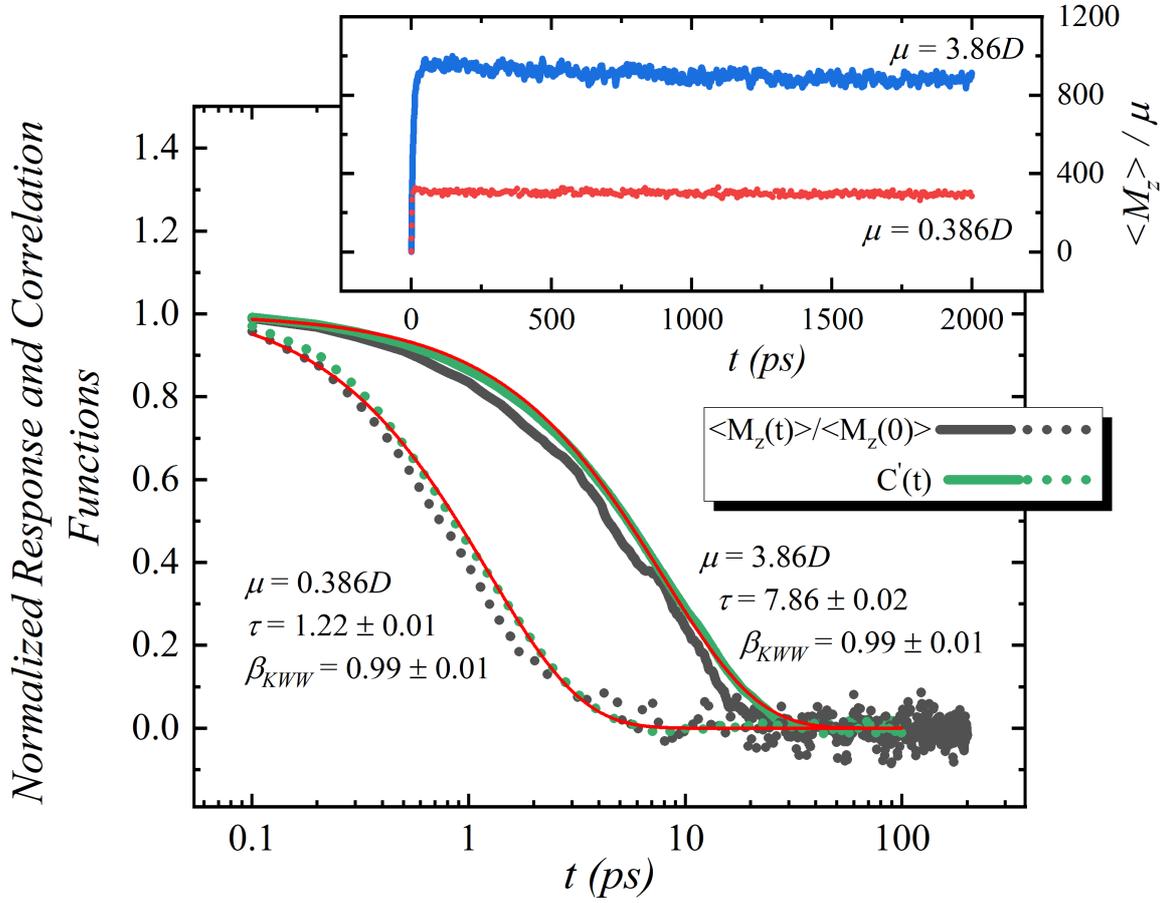

Fig.2 (color online)
The normalized response, $\langle M_z(t) \rangle / \langle M_z(0) \rangle$, and correlation, $C'(t)$, functions are presented. The black lines (solid and dotted) represent the response functions and are calculated basing on the results presented in the inset. The green lines (solid and dotted) are obtained from fluctuations at equilibrium conditions. The dotted lines depict results obtained for weakly polar liquid, whereas the solid ones for strongly polar liquid. The red lines represent fits of $C'(t)$ by a decay function described by Eq. (1). Inset shows the time evolution of the systems' total dipole moment in the z direction, which are forced by the applied external electric field. The blue line is obtained for strongly polar liquid, whereas red line for weakly polar one. Note that, if all molecules would be arranged parallelly to $E_z$, $\langle M_z \rangle / \mu = N$.

In the inset one might observe an average value of the dipole moment oriented in the direction z, $\langle M_z \rangle$, which in order to fairly compare changes taking place within systems has been scaled by $\mu$. Interestingly, already this part of our studies reveals evident differences between both RM systems. RM with higher $\mu$ value (blue line) exhibits higher orientational polarizability despite that corresponding $E_z$ have been applied to both systems. It suggests the existence of an additional contributions to $\boldsymbol{P}$, which might originate from a collective behavior of molecules. Another crucial observation is that both systems reach an equilibrium within time of $2ns$. Hence, the applied $E_z$ does not polarize the system anymore and the system's reaction onto an outer disturbance can be quantified during the second part of our experiment. When the external electric field is immediately removed, system responds, which is observed by a decrease in $\boldsymbol{P}(t)$. As we already noted, $\boldsymbol{P}(t)$ can be quantified using of $\boldsymbol{M}(t)$. However, in our case, only component in $z$ direction is of interest. Therefore, in the Fig.2 we present the normalized function of system's total dipole moment, $\langle M_z(t)\rangle/\langle M_z(0)\rangle$, the starting point is the time at which we turn off $E_z$. Also in this figure, one can observe evident differences between $\langle M_z(t)\rangle/\langle M_z(0)\rangle$ for both systems. The RM system with a higher $\mu$ value relaxes slower (solid line) than the one with a smaller $\mu$ value (dotted line). Since, our experiment is performed in the linear response regime, $\tau$ does not depend on $E_z$ magnitude. Moreover, in Fig.2 we depict normalized $C(t)$, $C'(t) = C(t)/\langle \boldsymbol{M}(0)\boldsymbol{M}(0)\rangle$, calculated basing on the third step of our experiment. It is crucial because according to the fluctuation dissipation theorem the response of system onto external disturbance is related to fluctuations at equilibrium [7], $C(t)$. As one can see, for both studied systems, $\langle M_z(t)\rangle/\langle M_z(0)\rangle$ indeed corresponds to the fluctuations of a corresponding quantity at equilibrium. The slight differences result from a limitation in $\langle M_z(t)\rangle/\langle M_z(0)\rangle$ averaging, which is done for 50 independent simulations, whereas $C'(t)$ is an average of 50 simulations and 1000 functions established from one simulation run. Correspondence of $\langle M_z(t)\rangle/\langle M_z(0)\rangle$ and $C'(t)$ is crucial for further studies because we can employ simulations at equilibrium, which significantly improves data statistics. At this point, it is also worth mentioning that a similar value of $\beta_{KWW}$ for both systems can be justified by the fact that we examine thermodynamic conditions, which are far from the glass transition. The dependence of the $C'(t)$ on thermodynamic conditions is discussed later.

Before, in Fig.3, we present calculated $C'(t)$ and its contributions from self- and cross-correlations, $C'_s(t) = C_s(t)/\langle M(0)M(0)\rangle$ and $C'_c(t) = C_c(t)/\langle M(0)M(0)\rangle$ respectively.

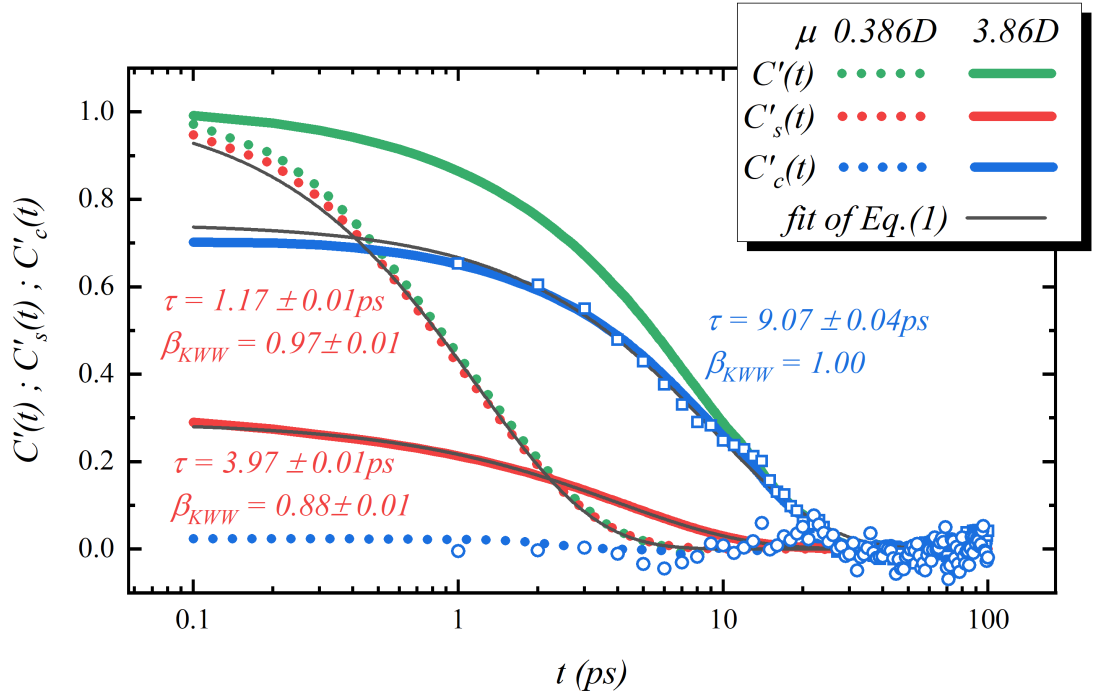

Fig.3 (color online)
The normalized total dipole correlation functions and their self- and cross-components are presented for two model systems. The dotted lines are calculated for a representative of weakly polar liquids, whereas solid lines are obtained for strongly polar liquid. The cross-correlations contributions are calculated according to the relationship $C'_c(t) = C'(t) - C'_s(t)$. The black line represents the fit of a decay function described by Eq. (1). The values of the stretching parameter are presented for self- and cross-correlations contributions to total dipole correlation function. The open symbols depict cross-correlation contributions calculated directly, i.e., according to Eq. (4). Due to the computational effort data are averaged over 50 independent functions (in contrast data presented as lines are averaged over 5000 independent functions).

Interestingly, even though $C'(t)$ exhibit similar shapes for both RM systems, the contributions to $C'(t)$ from the self- and cross-correlations are totally different. In the case of the weakly-polar system (dotted lines), $C(t)$ is entirely dominated by the self-correlations. The role of $C_c(t)$ is marginal because it gives input to $C(t)$ at the level of less than 5 percent. However, totally different situation takes place for the strongly-polar liquids (see results for the second RM system). The increase in $\mu$ value makes that, due to a stronger dipole-dipole interactions, the correlations between neighboring molecules become noticeable. It causes that the role of $C_c(t)$ in the total dipole correlation function increases. Consequently, we observe that $C_c(t)$ dominates over $C_s(t)$ for RM system with high $\mu$ value (solid lines). However, in this case the role of $C_s(t)$ cannot be neglected because it is still at the level of 30 percent. Hence, our findings proves that BDS spectra reflects mainly self-correlations only for the weakly polar liquids. Therefore, exclusively for those systems, the BDS and DDLS experiments could correspond to each other. We would like to put readers' attention as well that, similarly to previous suggestions, $C_s(t)$ vanishes faster than $C_c(t)$. It differs from the results obtained by Zhou and Bagchi [40] for self-consistent continuum model of Nee and Zwanzig [41], who find that within studied system a single particle orientational relaxation is slower from the collective orientational relaxation. The noted discrepancies might be caused by the absence of the translational motion in the examined dipolar lattice model. Nevertheless, our results are in accord with predictions of Déjardin *et al.* [17,18], who suggest that the collective mode relaxes slower than single molecular one.

As we already noted the role of self-correlations in the relaxation process cannot be negligible for strongly polar system. Therefore, the corresponding BDS spectrum contains self- and cross- contributions, wherein the self-correlations give a smaller input located at higher frequencies. The latter might lead to spectra broadening, see schematic representation in Fig.1, especially that our results reveal as well, that $C_c(t)$ exhibits narrower shape of the relaxation spectrum than $C_s(t)$. It directly reflects postulate by Pabst *et. al.*

Finally, in Fig.4 we present the evolution of $C'(t)$ and $C'_c(t)$ when the studied systems approaching to the glass transition, which have been realized by the isothermal compression.

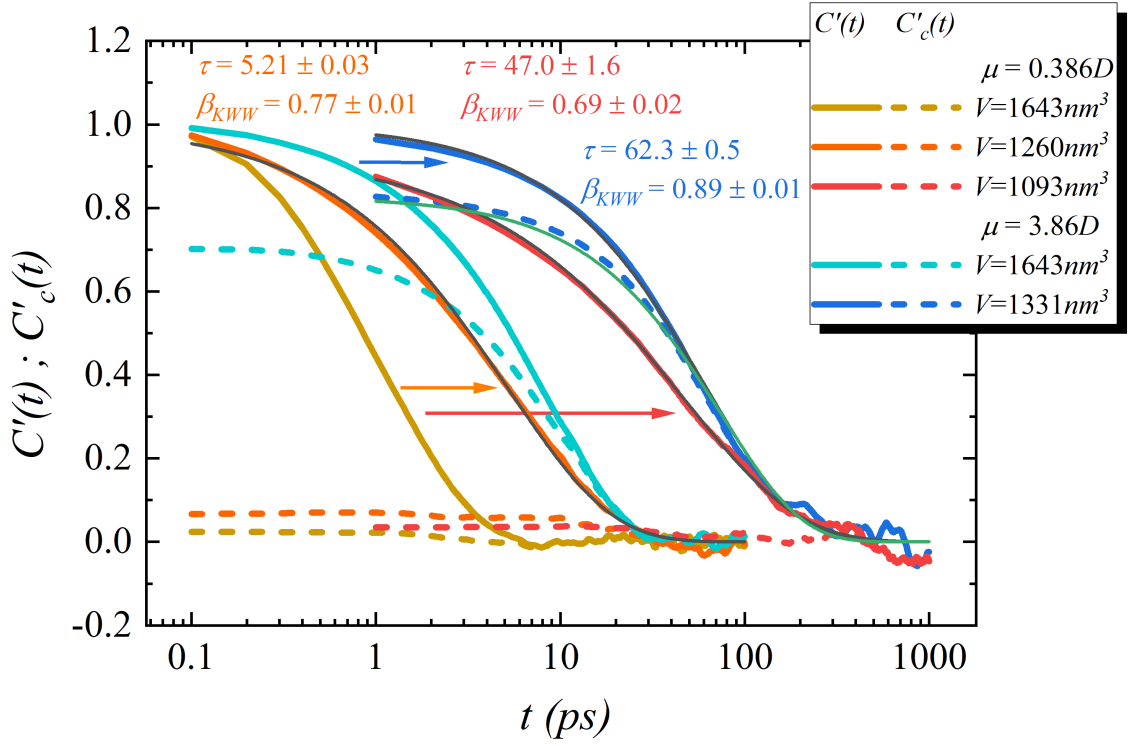

Fig.4 (color online)
The normalized total dipole correlation functions and their cross-components are presented for two studied model systems at different thermodynamic conditions. The arrows indicate the compression of weakly (orange and red) and strongly (blue) polar liquids. The black lines depict the fit of the function described by Eq. (1) to the obtained data. The green line is a fit of the Debye function to the cross-correlations for the compressed strongly polar system.

Comparing, $C'(t)$ characterized by the similar relaxation times, one can see that $C'(t)$ for weakly polar system is much broader. Furthermore, for this system, a slowing down in molecular dynamics results in an evident change in $\beta_{KWW}$ (from 0.99 to 0.77 and then to 0.69). Hence, one we might suspect that $\beta_{KWW}$ tends to the suggested value of 0.5. However, examination of the corresponding thermodynamic conditions is a very challenging task due to limitation of the computational experiment. Contrary, for strongly polar liquid, $C'(t)$ changes only slightly, i.e., $\beta_{KWW}$ of the corresponding functions changes from 0.99 to 0.89. Hence, the obtained $C'(t)$ follows not only the postulate by Pabst *et al.* but also the experimentally established correlation between the dipole moment value and the shape of the BDS spectrum. [4] The latter could be explained by the role of the cross-correlations in systems' relaxation process. For weakly polar system $C'_c(t)$ is marginal and persist on the same level for the majority of time. It makes that the behavior of $C'(t)$ is determined solely by the self-correlations, which are much broader. On the other hand, $C'_c(t)$ visibly dominates $C'(t)$ for strongly polar system. The decrease in the self-correlation is visible only at the initial stage of the relaxation process, which makes that its contribution is observed at frequencies higher than the dielectric loss peak. The latter results only in a tiny broadening of the spectrum because $C'_c(t)$ can be still described by the Debye-like behavior with a very good accuracy.

Summarizing, in this letter, we contribute to the recently proposed explanation of the differences between obtained by the various experimental methods, i.e., BDS and DDLS. Our experiment delivers clear evidence for the existence of the direct relationship between the broadening of the dielectric spectra and the role of cross-correlations of the molecules' dipoles moments. We present that the cross-correlations practically do not occur for weakly polar molecules and that for this system the corresponding spectrum is distinctly widened. Contrary, the total-dipole correlations function is almost entirely dominated by the cross-term for strongly polar liquids. Since the cross-correlations are characterized by the less broad relaxation function, the corresponding dielectric spectrum is only slightly widened. In this case self-correlations give a tiny contribution at the initial stage of the dielectric relaxation process. It implies that, the established in the literature the relationship between the dipole moment value and the BDS spectral shape can be immediately justified. Therefore, our findings not only strongly support the postulate by Pabst *et al.,* but also proves the existence of the experimentally observed correlation. Consequently, our paper gives the cornerstone for the proper interpretation of the nature of BDS spectrum.

**ACKNOWLEDGMENTS**

The authors are deeply grateful for the financial sup- port by the Polish National Science Centre within the framework of the Maestro10 project (Grant No. UMO- 2018/30/A/ST3/00323).